# Towards angiosperms genome evolution in time

Serge N. Sheremet'ev[1] and Yuri V. Gamalei

*Department of Ecological Physiology, Komarov Botanical Institute, Russian Academy of Sciences, St. Petersburg 197376, Russia*

**Abstract** In this communication, direction of evolutionary variability of parameters of genome size and structurally functional activity of plants in angiosperm taxa among life forms, are analyzed. It is shown that, in the Cretaceous–Cenozoic era, the nuclear genome of the plants tended to increase. Functional genome efficiency (intensity of functions per pg of DNA) decreased as much as possible from highest, at trees and lianas of rain and monsoonal forests of the Paleogene, to minimum, at shrubs, perennial and annual grasses of meadow–steppe vegetation appeared in the Neogene. Environmental changes in temperature, humidity and $CO_2$ concentration in adverse direction, critical for vegetation, are discussed as the cause of growth of evolutionary genome size and loss in its functional efficiency. The growth of the genome in the Cenozoic did not lead to the intensification of functions, but rather led to the expansion of the adaptive capacity of species. Growth of nuclear DNA content can be considered as one of the effective tools of an adaptogenesis.

## INTRODUCTION

Nuclear DNA content (nDNAc) in plant and animal cells is being investigated for more than 50 years (Bennett and Leitch, 2005a); and so far, the nDNAc has been defined for 6.5 thousand angiosperm species. This information is summarized in tabular forms in several publications (Bennett, 1972; Bennett and Smith, 1976, 1991; Bennett et al., 1982, 1998, 2000; Bennett and Leitch, 1995, 1997, 2005a; Hanson et al., 2001a, b; Hanson et al., 2003, 2005; Suda et al., 2003; Zonneveld et al., 2005) and in free online database (Bennett and Leitch, 2005b). We enlarged this database by adding more information about geological age of plant genera (after Daghlian, 1981; Muller, 1981; Zavada and Benson, 1987; Benton, 1993; Collinson et al., 1993; Herendeen and Crane, 1995; Wing and Boucher, 1998; Crepet et al., 2004; Song Zhi–Chen et al., 2004; Martínez–Millán, 2010; Grimsson et al., 2011 and The Paleobiology Database (PBDB – http://flatpebble.nceas.ucsb.edu/cgi–bin/bridge.pl)) and their growth forms.

This research is undertaken to detect changes in nDNAc, number of chromosomes, ploidy levels in angiosperms and their taxonomical subunits and different growth forms, during the course of evolution from late Cretaceous to Plio–Pleistocene. Influence on dynamics of the climatic factors as well synchronism with changes of structurally functional characteristics of plants is studied. On possibility of such coordination point the known facts of influence on a wide range of properties of plant nDNAc is documented (Grime, 1998; Reeves et al., 1998; Prokopowich et al., 2003; Jovtcheva et al., 2006; Beaulieu et al., 2007a, b; Beaulieu et al., 2008; Knight and Beaulieu, 2008 и др.).

---

[1] Corresponding author. E-mail: Serge Sheremet'ev, sn.sheremetiev@gmail.com

Being engaged in this work, we understood nevertheless some vagueness of such constructions about the assumption what data obtained from extant species has remained unchanged over geological time. This assumption is quite vulnerable relative to ploidy levels, to a lesser extent – relative to the numbers of chromosomes and the amount of DNA content in haploid sets of chromosomes. However this question, in our opinion, is interesting and demands the preliminary analysis and discussion at least. We offer the certain probabilistic approach which correctness can be estimated by the general logic of angiosperms evolution, and also by coincidence of the received curves with climatic data. However, it should be noted that there are different approaches to this problem (Masterson, 1994; Franks et al., 2012). We dare to hope that in the future, these approaches would supplement each other.

## DATA SETS

In addition to above mentioned data, following data sets were also used: cell cycle time (Francis et al., 2008), leaf functional traits $A_{max}$ (photosynthetic capacity, per leaf dry mass) and SLA (specific leaf area, per leaf dry mass) (Wright et al., 2004), leaf vein density (Brodribb and Field, 2010; Feild et al., 2011), plant water relations (index of plant water relations complexity and partial volume of intercellular spaces) (Sheremet'ev, 2005), and chlorophyll content (per leaf dry mass) (Lubimenko, 1916), oxygen isotope ratio in shells of planktonic foraminifera and brachiopods from J. Veizer's database (http://www.science.uottawa.ca/geology/isotope_data/) (Veizer et al., 1999), arid areas (computed after paleomaps by Scotese, 2003; Akhmetiev, 2004; Chumakov, 2004a). All data were averaged by epochs (Ogg et al., 2008).

## TERMINOLOGY

The labeling of nDNAc is done by the use of the term "C–value" with various prefixes – 1C, 2C, 3C etc. It was used for the first time by Swift (Swift, 1950) without any definition. Later Bennett and Smith (1976) pointed out, that Swift in personal communication informed that the letter C stands for 'constant', i.e. the amount of DNA characteristic of a particular genotype. They defined the C–value (or 1C value) for any genotype as the DNA content of the unreplicated haploid chromosome complement (Bennett and Smith, 1976). Similarly, the measure of DNA content of unreplicated non–reduced (zygotic, diplophasic) complements will be 2C (irrespective to ploidy level) (Greilhuber et al., 2005).

Unlike the term "C–value" the term "genome size" is often used to designate the quantity of DNA in meiotic reduced diploid set of chromosomes or monoploid one (meaning the polyploid set contains more that one genome) (Bennett et al., 1998). Consistent use of the term "genome size" in a narrow sense is often impossible due to uncertainty of the degree of ploidy. Therefore many authors prefer to use this term, as well as "C–value" in a broader sense irrespective of the ploidy level. In this sense, the terms "C–value" (with prefix 1) and "genome size" are synonymous (Greilhuber et al., 2005).

The complete chromosome set with number n (reduced) irrespective of the degree of generative polyploidy was offered the term "holoploid genome" (Greilhuber et al., 2005; Greilhuber and Doležel, 2009). It was proposed to use the terms "holoploid genome size" and "C–value" (or 1C) to determine size of genome (Greilhuber et al., 2005). In that case 2C will be related to an unreplicated, unreduced chromosome set (2n) regardless the degree of ploidy and will characterize total amount of the nuclear DNA (in diplophase), and the genome size will correspond to the term "holoploid genome size" (or 1C). In this article, we will hold these definitions for these terms.

## VARIABILITY OF TRAITS

Traits that characterize nuclear DNA of angiosperms (2C, the number of chromosomes and ploidy level) vary at broad limits. Especially it concerns the nDNAc whose coefficient of variation is more than 150 % (Table 1). More moderate variations are discovered in chromosome numbers (71%) and ploidy levels (59%). The wide variation of genome size in flowering plants (from 0.05 to 140 pg) is a part of greater variety of 1C among eukaryotic organisms – from 0.009 to 700 pg (see, for example: Leitch et al., 1998; Gregory, 2005; Patrushev and Minkevich, 2006, 2007). This variation, which is not related with the taxonomic positions of the species and their phenotypic complexity, was called the "C–value paradox" (Thomas, 1971).

However, the fractile analysis shows that the variability of the data is not so considerable (Table 2). The difference of the nDNAc between the top and bottom deciles (fractiles 0.9 and 0.1 respectively) makes only 32.3 pg. In other words, the vast majority of array values (80%) are in limits of 1.2–33.5 pg. By such consideration the genome sizes are limited at a range of 0.6–16.7 pg; when, number of the chromosomes changes from 14 to 48, the ploidy levels do not exceed 4 (Table 2). Comparison of these data with the variability which is shown in

**Table 1.** Nuclear DNA content (2C, pg), number of chromosomes (2n) and ploidy levels (PL) in different growth forms of angiosperms

| Statistics | Angiosperms | | | Monocots | | | | Dicots | | | |
|---|---|---|---|---|---|---|---|---|---|---|---|
| | All growth forms | Trees and shrubs | Herbs | Lianas | Herbs | Palms | Lianas | Trees | Shrubs | Herbs | Lianas |
| **2C** | | | | | | | | | | | |
| n | 6949 | 1133 | 5001 | 68 | 2564 | 107 | 23 | 359 | 253 | 2437 | 45 |
| X | 13.6 | 4.4 | 15.9 | 7.9 | 23.1 | 8.4 | 10.6 | 3.1 | 5.6 | 8.3 | 6.5 |
| Sx | 21.0 | 11.1 | 23.1 | 8.5 | 29.2 | 10.7 | 6.7 | 4.3 | 21.0 | 8.9 | 9.0 |
| CI$_X$ | 0.5 | 0.6 | 0.6 | 2.0 | 1.1 | 2.0 | 2.7 | 0.4 | 2.6 | 0.4 | 2.6 |
| Me | 6.2 | 2.1 | 8.5 | 4.0 | 12.8 | 5.8 | 10.0 | 1.9 | 2.2 | 5.3 | 1.1 |
| CI$_{Me}$ | 1.4 | 1.9 | 1.8 | 5.7 | 3.2 | 5.8 | 7.8 | 1.3 | 7.4 | 1.0 | 7.5 |
| V | 154.2 | 255.9 | 144.9 | 106.9 | 126.3 | 127.1 | 62.7 | 137.1 | 378.2 | 106.9 | 137.8 |
| CI$_V$ | 5.9 | 24.0 | 6.5 | 41.0 | 7.9 | 38.9 | 41.4 | 22.9 | 75.2 | 6.8 | 65.0 |
| Min | 0.1 | 0.9 | 0.1 | 0.6 | 0.3 | 0.5 | 0.8 | 0.3 | 0.2 | 0.1 | 0.6 |
| Max | 280 | 105.8 | 280 | 31.6 | 280.0 | 78.2 | 22.3 | 32 | 181.5 | 63.4 | 31.6 |
| **2n** | | | | | | | | | | | |
| n | 5908 | 979 | 4286 | 44 | 2084 | 80 | 17 | 310 | 219 | 2202 | 27 |
| X | 27.7 | 33.0 | 25.8 | 35.5 | 27.9 | 38.9 | 36.7 | 29.0 | 33.2 | 23.7 | 34.7 |
| Sx | 19.6 | 22.3 | 16.6 | 6.9 | 17.0 | 63.2 | 6.4 | 17.6 | 20.5 | 15.9 | 7.2 |
| CI$_X$ | 0.5 | 1.4 | 0.5 | 2.0 | 0.7 | 13.8 | 3.1 | 2.0 | 2.7 | 0.7 | 2.7 |
| Me | 22 | 26.0 | 20.0 | 38.0 | 24.0 | 32.0 | 36.0 | 24 | 30 | 18.0 | 38 |
| CI$_{Me}$ | 1.4 | 4.0 | 1.4 | 5.8 | 2.1 | 39.6 | 8.8 | 5.6 | 7.8 | 1.9 | 7.7 |
| V | 70.8 | 67.4 | 64.3 | 19.4 | 61.0 | 162.3 | 17.5 | 60.7 | 61.6 | 66.9 | 20.6 |
| CI$_V$ | 2.9 | 6.8 | 3.1 | 9.2 | 4.2 | 57.4 | 13.5 | 10.9 | 13.2 | 4.5 | 12.5 |
| Min | 4 | 8.0 | 4 | 16 | 6 | 22 | 30 | 14 | 8 | 4 | 16.0 |
| Max | 596 | 196.0 | 144 | 50 | 132 | 596 | 50 | 196 | 174 | 144.0 | 38.0 |
| **PL** | | | | | | | | | | | |
| n | 5708 | 912 | 4165 | 39 | 1957 | 80 | 11 | 272 | 209 | 2208 | 28 |
| X | 2.8 | 2.6 | 2.8 | 2.0 | 3.0 | 2.0 | 2.0 | 2.4 | 2.9 | 2.6 | 2.0 |
| Sx | 1.6 | 1.6 | 1.7 | | 1.7 | | | 2.0 | 1.5 | 1.6 | |
| CI$_X$ | 0.0 | 0.1 | 0.1 | | 0.1 | | | 0.2 | 0.2 | 0.1 | |
| Me | 2 | 2.0 | 2.0 | 2.0 | 2.0 | 2.0 | 2.0 | 2 | 2 | 2.0 | 2 |
| CI$_{Me}$ | 0.1 | 0.3 | 0.1 | | 0.2 | | | 0.7 | 0.6 | 0.2 | |
| V | 58.8 | 62.3 | 58.9 | 0.0 | 57.2 | 0.0 | 0.0 | 84.0 | 52.1 | 59.6 | 0.0 |
| CI$_V$ | 2.5 | 6.5 | 2.9 | | 4.1 | | | 16.1 | 11.4 | 4.0 | |
| Min | 2 | 2.0 | 2 | 2 | 2 | 2 | 2 | 2 | 2 | 2 | 2.0 |
| Max | 24 | 24.0 | 20 | 2 | 19 | 2 | 2 | 24 | 8 | 20.0 | 2.0 |

Here and in Tables 2–3 are shown the results of our processing database "C–Value" (Bennett, Leitch, 2005b); here and in Table 3: n – number of measurements; X – mean; Sx – standard deviation; Me – median; V – coefficients of variation; Min, Max – minimum and maximum values; CI$_X$, CI$_{Me}$, CI$_V$ – confidence intervals for the mean, median and coefficient of variation.

**Table 2.** Fractile analysis of the series distribution of genome size (1C, pg), nuclear DNA content (2C, pg), numbers of chromosomes (2n) and ploidy levels (PL)

| Fractile | 1C | 2C | 2n | PL |
|---|---|---|---|---|
| 0.1 | 0.6 | 1.2 | 14 | 2 |
| 0.25 | 1.2 | 2.4 | 16 | 2 |
| 0.5 | 3.1 | 6.2 | 22 | 2 |
| 0.75 | 8.2 | 16.4 | 34 | 3 |
| 0.9 | 16.7 | 33.5 | 48 | 4 |

Table 1 demonstrates that the general opinion on excessive variability of the nDNAc is somewhat exaggerated.

In this regard the question of intrageneric variability of the traits is of interest (we investigated data only for genera for which the number of observations was not less than three). Average intrageneric coefficients of variation of the genome sizes and the nDNAc are 36%. Coefficients of variation scattering were in a wide range (0–165 %). However the top decile was 62%, i.e. 90% of coefficients of variation were lower than 62%. Average coefficient of variation for genera chromosome numbers was 24% (top decile – 58%), and for ploidy levels – 21% (top decile – 51%). Therefore we suggest that intrageneric variability of the traits was high, but not extremal. These properties were observed due to right–tailed distribution of the data.

All these arrays have the extremely asymmetric (right–tailed) distributions (unfortunately, it was impossible to fit well these distributions). However, for all consideration, there was no obstacle for the use of parametric criteria. It is established (Hill and Lewicki, 2007) that as the sample size increases, the shape of the sampling distribution (i.e., distribution of a statistic from the sample) approaches to the normal, even if the distribution of the variable in question is not normal. However, as the sample size ($n \gg 30$) (of samples used to create the sampling distribution of the mean) increases, the shape of the sampling distribution becomes normal. Many studies have shown that the consequences of violations of the assumption of normal distribution are less severe than previously thought (Hill and Lewicki, 2007).

Thus, typical changes (within fractiles 0.1–0.9) of the considered traits are contained in not so wide limits. In addition, we can not say that the size of genomes and nDNAc does not depend on the taxonomic status of plants and their phenotypic complexity.

As reported earlier (Bharathan et al., 1994; Hanson et al., 2003), mean value of nDNAc of dicots was about three times less than the monocots (Table 3) ($p < 0.001$) It was shown earlier that the ancestral monocot groups (as well as the ancestral groups of dicotyledons) had the small genome sizes, in contrast to the derivative taxa those had very high content of nDNAc (Leitch et al., 1998; Soltis et al., 2003; Leitch et al., 2005). This view is confirmed by quantitative analysis of the data.

**Table 3.** Nuclear DNA content (2C, pg), number of chromosomes (2n), and ploidy levels (PL) in monocots and dicots

| Statistics | Monocots | | | Dicots | | |
|---|---|---|---|---|---|---|
| | 2C | 2n | PL | 2C | 2n | PL |
| n | 2927 | 2373 | 2236 | 4022 | 3535 | 3472 |
| X | 22.7 | 28.6 | 2.9 | 7.0 | 27.1 | 2.6 |
| Sx | 27.9 | 21.3 | 1.7 | 9.6 | 18.3 | 1.6 |
| $CI_X$ | 1.0 | 0.9 | 0.1 | 0.3 | 0.6 | 0.1 |
| Me | 13.3 | 24.0 | 2.0 | 3.5 | 22.0 | 2.0 |
| $CI_{Me}$ | 2.9 | 2.5 | 0.2 | 0.8 | 1.7 | 0.1 |
| V | 122.9 | 74.5 | 56.9 | 136.7 | 67.8 | 59.7 |
| $CI_V$ | 7.2 | 4.8 | 3.8 | 6.8 | 3.6 | 3.2 |
| Min | 0.3 | 4.0 | 2.0 | 0.1 | 4.0 | 2.0 |
| Max | 280.0 | 596.0 | 19.0 | 181.5 | 196.0 | 24.0 |

Evolutionary progress was defined by morphological evolution and, in particular, by changing of existing and emergence of new forms of growth (palms, lianas, herbs). The point of view on the evolution of growth forms from trees and shrubs in the direction of herbs is shared by many authors (see Takhtajan, 2009). In a series of growth forms from trees to herbs with a tendency to increase of nDNAc can be traced (Fig. 1A). However, this tendency isn't unidirectional, as it can be seen in the picture on the plane (Fig. 1A), but most likely characterized by numerous branchings (Fig. 1B) and parallel ways.

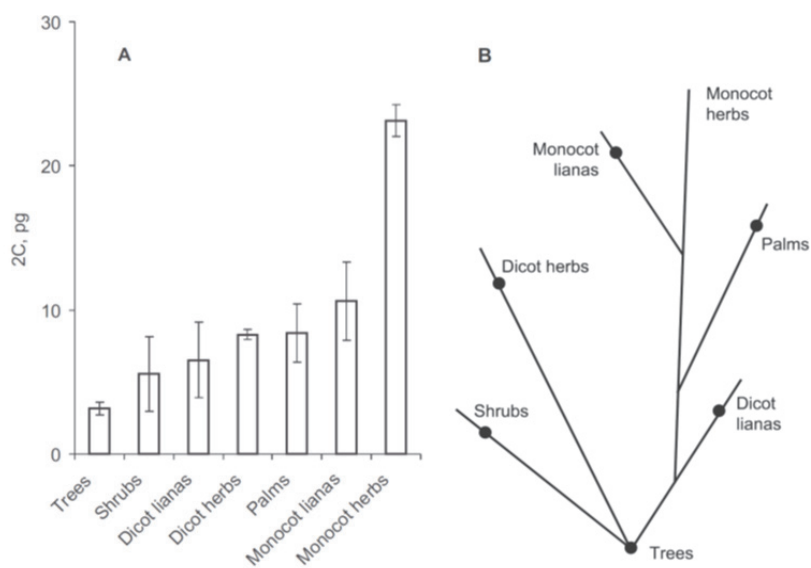

Fig. 1. Changes of nuclear DNA content in the array of growth forms (A) and possible ways of transformation of growth forms of angiosperms (B).

## CHANGES OF TRAITS IN LATE CRETACEOUS–CENOZOIC

Determination of geologic age of angiosperm genera allowed to trace the changes of nDNAc, chromosome numbers, and ploidy levels in late Cretaceous–Cenozoic and compared these time series with important characters of paleoclimates.

### NUCLEAR DNA CONTENT

The nDNAc of angiosperms, as already noted, had a tendency to increase during the course of biological evolution. A more detailed study of this trend showed that the minimum value of nDNAc was observed in the initial period of the evolution of angiosperms – in Early and Late Cretaceous (Fig. 2A). In the Paleocene there was some increase in nDNAc, followed by a decrease in the Eocene. Rate of angiosperms evolution (as number of families appeared in some geological epoch divided by duration of this epoch) changed in an antiphase. In other words, the rate of appearing of new families of angiosperms increased in geological epochs when nDNAc decreased. On the contrary the increase of nDNAc was accompanied by a decrease in the rate of evolution.

A slow decrease in temperature after the early Eocene climatic optimum was followed by a drastic climatic cooling and large–scale glaciation of the Antarctica (Kennett 1977; Lear et al., 2000; Zachos et al., 2001; DeConto and Pollard 2003a, b; Pollard and DeConto 2003, 2005). Sometimes later, glaciation spread over Greenland (later Eocene–early Oligocene: Eldrett et al., 2007). After that, symmetric glaciation of both the poles occurred (Tripati et al., 2005; Moran et al., 2006), and the temperature gradient between the equatorial zone and the poles increased (Nikolaev et al., 1998). It resulted in beginning of transition from the ''warm biosphere'' to the ''cold biosphere'' (Akhmetiev, 2004).

In these conditions of Oligocene glaciation, a steep increase in nDNAc and a steep decrease in the rate of angiosperms evolution occurred (Fig. 2A). Then, starting from the Miocene, the nDNAc gradually decreased, with the onset of the rise of the rate of angiosperm evolution.

In monocots and dicots the changes of nDNAc in Cretaceous–Eocene occurred more or less in parallel (Fig. 2B). However in Oligocene, reactions of plants to climatic changes were opposite. As a response to cooling and reduction of carbon dioxide in the atmosphere in Oligocene, the nDNAc in dicots dramatically increased, and by contrast, sharply declined in monocots. Reaction of these angiospermic groups to a relative warming in Miocene remained opposite. It is not possible to find an explanation for this observation at this stage of research.

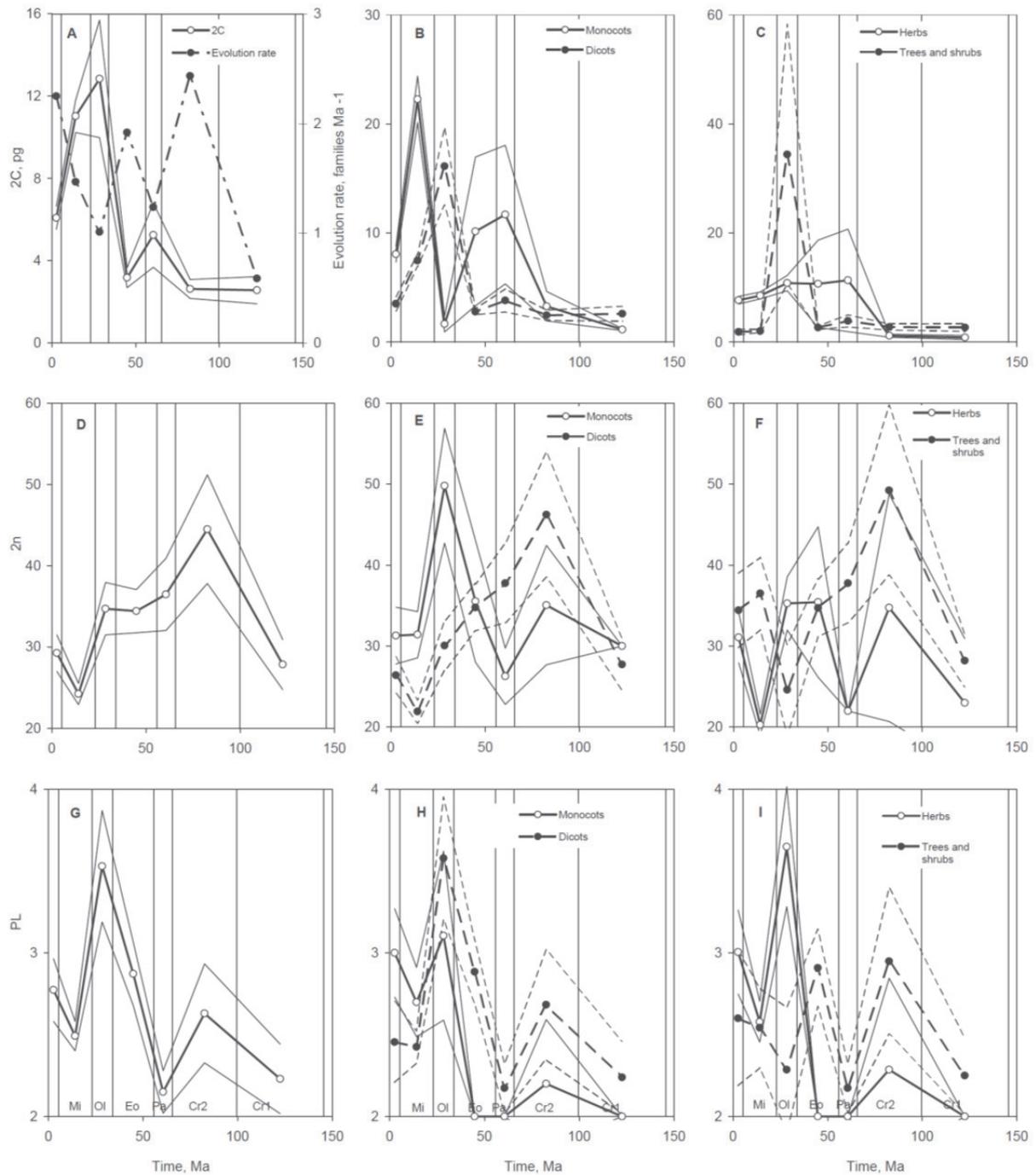

Fig. 2. Changes the rate of angiosperms evolution (A), nuclear DNA content (A, B, C), numbers of chromosomes (D, E, F) and ploidy levels (G, H, I) in angiosperms (A, D, G), in monocots and dicots (B, E, H), in herbs, trees and shrubs (C, F, I) in the Cretaceous-Cenozoic.

The data are grouped by epochs of the stratigraphic time scale (Ogg et al., 2008). Here and in figures 5–10: Mi – Miocene, Ol – Oligocene, Eo – Eocene, Pa – Paleocene, Cr2 – Late Cretaceous, Cr1 – Early Cretaceous. Confidence intervals are joined by thin lines without markers.

Each of growth forms in the ecological relation must be more homogeneous than the each taxon of the angiosperms. Nevertheless changes in nDNAc over the time in dicots (Fig. 2B) are identical to such changes in trees and shrubs (Fig. 2C) with the difference that amplitude of the plots is approximately twice higher in the woody plants. Herbs in Cretaceous–Eocene (Fig. 2C) approximately repeated the course of changes of nDNAc in monocots (Fig. 2B). However since Eocene the nDNAc of herbs changed slightly.

NUMBERS OF CHROMOSOMES

Number of chromosomes (2n) in most cases (90 %) is in not so widely limited from above 48 chromosomes (Table 2). It finds reflection on a curve of changes of 2n in angiosperms in time (Fig. 2D). In the angiosperms (on average) very large 2n at the initial stage of evolution (Early Cretaceous) was not there. However, in the Late Cretaceous the 2n had reached the maximum value in the history of this group of the plants (Fig. 2D) and was coincided with very high rate of evolution (Fig. 2A). The subsequent history of angiosperms showed a gradual reduction in the number of chromosomes in cell nuclei. The absolute minimum was passed in Miocene. In Pliocene–Pleistocene, approximately the same value was restored with which the history of the angiosperms began.

During the evolution of dicots, changes in 2n (Fig. 2E) were similar to the changes in 2n of trees and shrubs (Fig. 2F). In monocotyledonous plants (Fig. 2E) and dicotyledonous herbs (Fig. 2F) similar dynamics of 2n were observed in the Cretaceous–Cenozoic, except for a significant peak of this feature in monocots in the Oligocene.

RELATIONS BETWEEN 2N AND NDNAC

Under change over the entire range of nDNAc values, 2n on average remained unchanged or had a very weak tendency to decrease (Fig. 3A). It only means that under changes of nDNAc values the character of packaging of chromosomes should change. In more massive genomes the chromosomes should be more massive and vice versa. It confirms dependence of the ratio 2C/2n (specific chromosome mass – SCM) from 2C (Fig. 3B). On the other hand, the specific chromosomes number (2n/2C; SCN) decreased with increasing nDNAc (Fig. 3C). It can be deduced that the chromosomes in the less massive genomes became relatively larger in number, but lost mass; so conversely, in massive genomes, the chromosomes should be the more massive but in a relatively smaller number.

The study of dependence of cell cycle time (CCT) (data set by Francis et al., 2008) from SCM and SCN showed that the increase of SCM leads to an increase of CCT (Fig. 4A). In contrast, the increase of SCN is accompanied by a decrease of CCT (Fig. 4B).

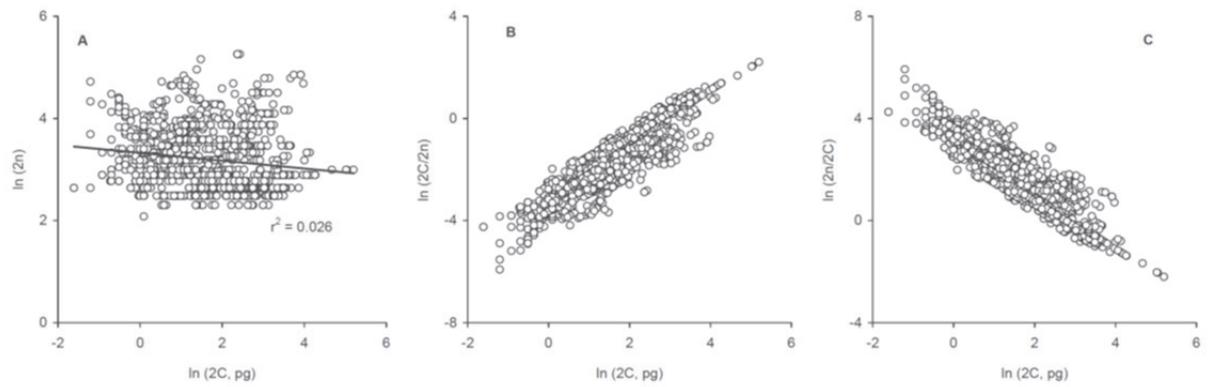

Fig. 3. Relation between nuclear DNA content and the number of chromosomes (A), specific chromosome mass (SCM) (B) and specific chromosomes number (SCN) (C) vs. nuclear DNA content.

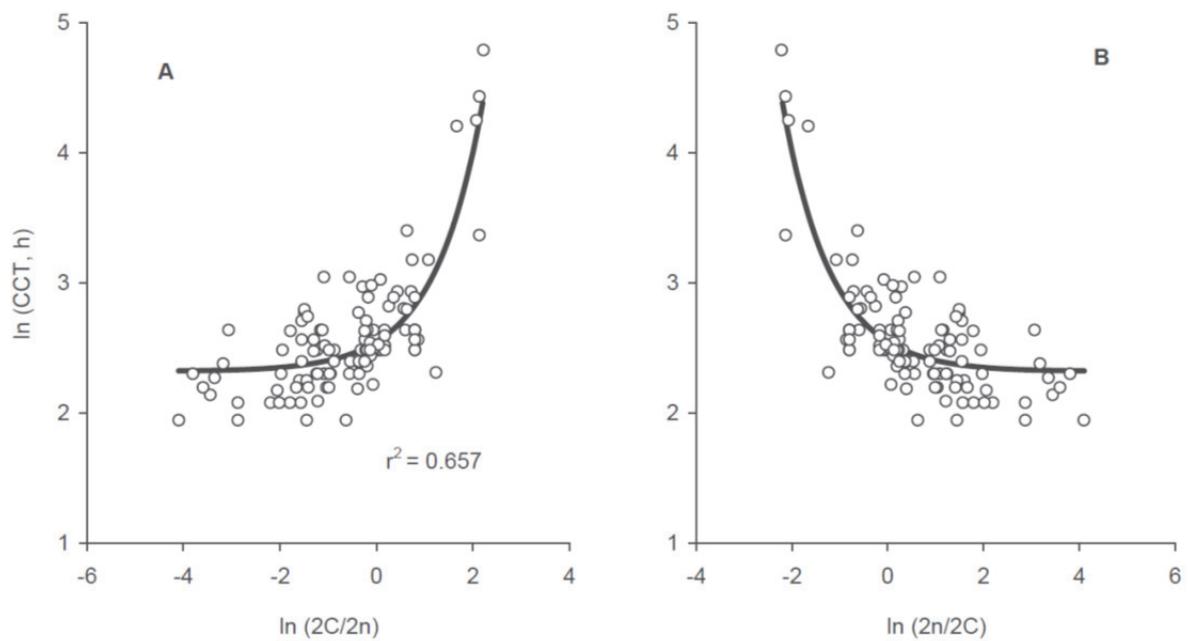

Fig. 4. Relations between cell cycle time (CCT) and specific chromosome mass (SCM) (A) and specific chromosomes number (SCN) (B).

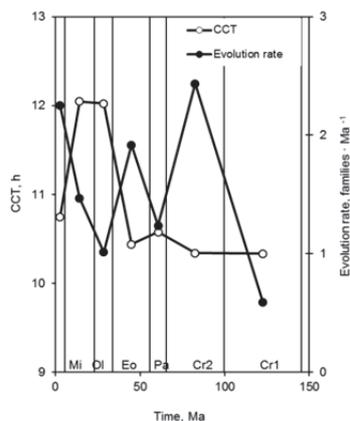

Fig. 5. The cell cycle time (CCT) and the evolution rate of angiosperms vs. time.

But to all appearances (see Fig. 3) the molecular machine during the cell cycle faster operates with a relatively large number of small chromosomes than with a small number of relatively large ones.

During the evolution of angiosperms there was a tendency to increase the nDNAc (Fig. 2A) and to reduce the number of chromosomes in the nucleus (Fig. 2D). Therefore, the evolution of flowering plants was accompanied by a trend of increasing CCT (Fig. 5). The maximum values of the CCT were achieved in the Oligocene and Miocene coinciding with the significant decrease in the rate of evolution of angiosperms (Fig. 5).

### PLOIDY LEVELS

Polyploidy is widespread among the angiosperms (Wendel and Doyle, 2005; Cui et al., 2006; Chen, 2007; Doyle et al., 2008; Soltis and Soltis, 2009; Jackson and Chen, 2010; Doyle and Egan, 2010). An important and integral part of the speciation and evolution of the flowering plants is an interspecific hybridization based on the duplication of the genome (De Bodt et al., 2005; Hegarty and Hiscock, 2005; Baack and Rieseberg, 2007; Mallet, 2007, 2008; Rieseberg and Willis, 2007; Soltis and Soltis, 2009; Wood et al., 2009; Whitney et al., 2010). According to some authors, it has become increasingly clear that the genome duplication events were associated with the most important steps in the evolution of angiosperms, such as the origin and early divergence, the evolution of the flower or the overcoming of the Cretaceous–Cenozoic boundary (Van de Peer et al., 2009a, b).

The study of database "C–Value" (Bennett and Leitch, 2005b) showed that the ploidy levels (PL) of most of the species does not exceed 4 (Table 2). Presumably the fluctuations of PL (on average) in the evolution did not exceed this limit (Fig. 2G–I). However, there was trend of some increase of PL during of the angiosperms evolution. The exceptions are woody plants (trees and shrubs), in them variations of PL occurred near the one level (without any trends) (Fig. 2I).

### CLIMATIC CONTROL OF DYNAMICS OF THE TRAITS

#### NUCLEAR DNA CONTENT

The important climatic factors such as global temperature and the activity of the hydrological cycle appeared to have a significant influence on nDNAc of the angiosperms (Fig. 6A, B). The nDNAc of the angiosperms remained rather low during the warmer geological epochs (Cretaceous–Eocene) (Fig. 6A). Global cooling in the Oligocene coincided with a significant increase of the nDNAc (Fig. 6A). In addition, the narrowing of the arid areas was accompanied by a relative decrease of the nDNAc (Fig. 6B). Climate aridization in the Oligocene (as well as global cooling) coincided with an increase of the nDNAc (Fig. 6B).

#### NUMBERS OF CHROMOSOMES

The numbers of chromosomes in all plants represented in the database "C–Value" (Bennett and Leitch, 2005b) have decreased since the Late Cretaceous (Fig. 2D). This

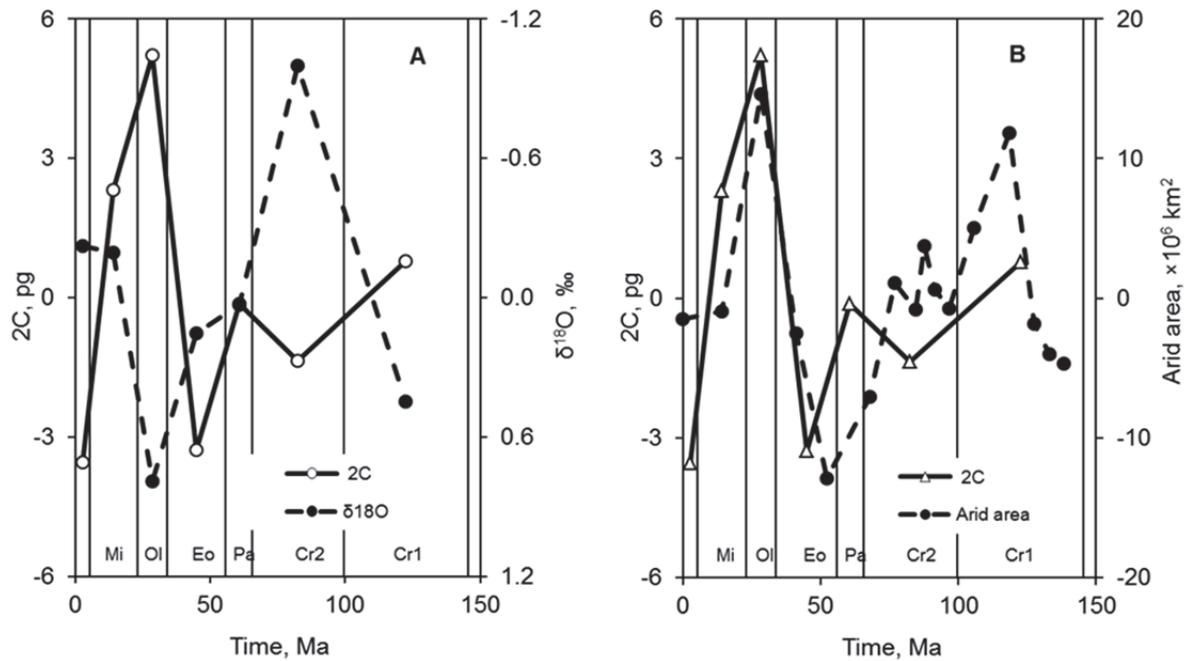

Fig. 6. The nuclear DNA content and the climate (A – $\delta^{18}$O; B – area of arid territories) vs. time.

Data detrended. $\delta^{18}$O – oxygen isotope ratio in shells of planktonic foraminifera and brachiopods, data averaging by epochs (Ogg et al., 2008) from J. Veizer's database (http://www.science.uottawa.ca/geology/isotope_data/) (Veizer et al., 1999). Arid area (here and in figures 7B, 8A-B white points – computed after maps by Scotese 2003, black points – after maps by Akhmetiev 2004 and by Chumakov 2004a, b, zero point – contemporary hyperarid, arid, and semi-arid areas of ice-free land after Middleton and Thomas 1997).

dynamics of 2n is almost identical to the dynamics of the decline in the global temperature (Fig. 7A). Similarly, the SCN was also decreased (i.e., the number of chromosomes per pg).

The global humidization of climate in Cretaceous–Paleocene was accompanied by a decrease of 2n in monocots (i.e., predominantly in herbaceous plants) (Fig. 7B). On the contrary, the arid climate in the Oligocene was related with the significant increase of 2n in this group of plants (Fig. 7B).

Dicots (Fig. 7C) (and also trees and shrubs) show the same picture of dynamics of 2n in Cretaceous–Cenozoic, as well as angiosperms as a whole (Fig. 7A) – in connection with changes in the global temperature.

### PLOIDY LEVELS

The PL in the angiosperms (as well as in the monocots and dicots) (Fig. 8A) and in the herbaceous plants (Fig. 8B) was varied in the geological past in accordance with the changes of climate aridity. The PL went down during humid epochs (Late Cretaceous–Eocene).

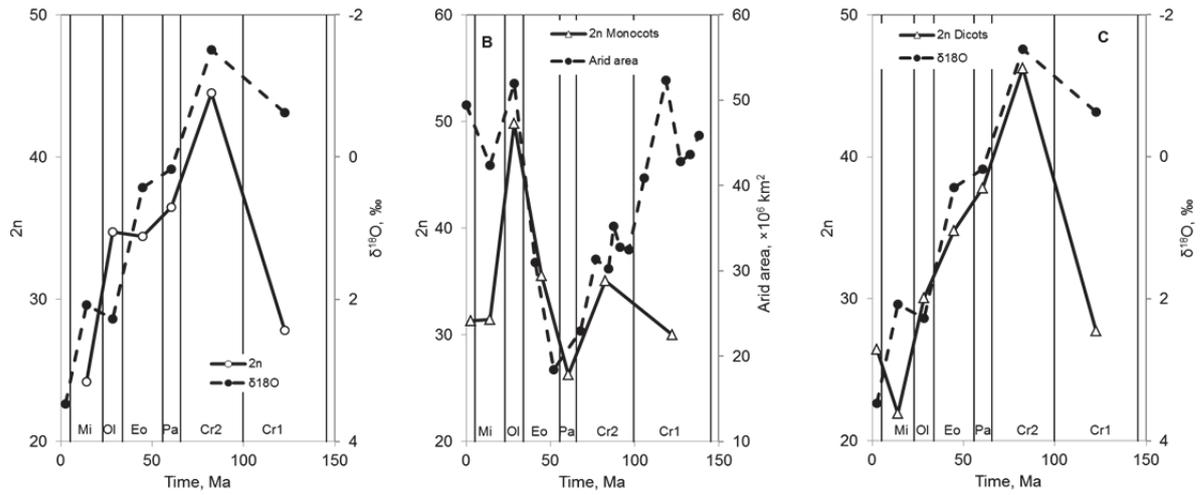

Fig. 7. The chromosome numbers and the climate (A, C – $\delta^{18}O$; B – area of arid territories) vs. time.

Increased PL in the Oligocene was associated with an aridization of the climate. The PL of the trees and shrubs was changed due to fluctuations in the global temperature (Fig. 8C). The PL in the woody plants was increased in the relatively warm epochs (Late Cretaceous–Eocene), and was declined in a relatively cold epoch (Oligocene).

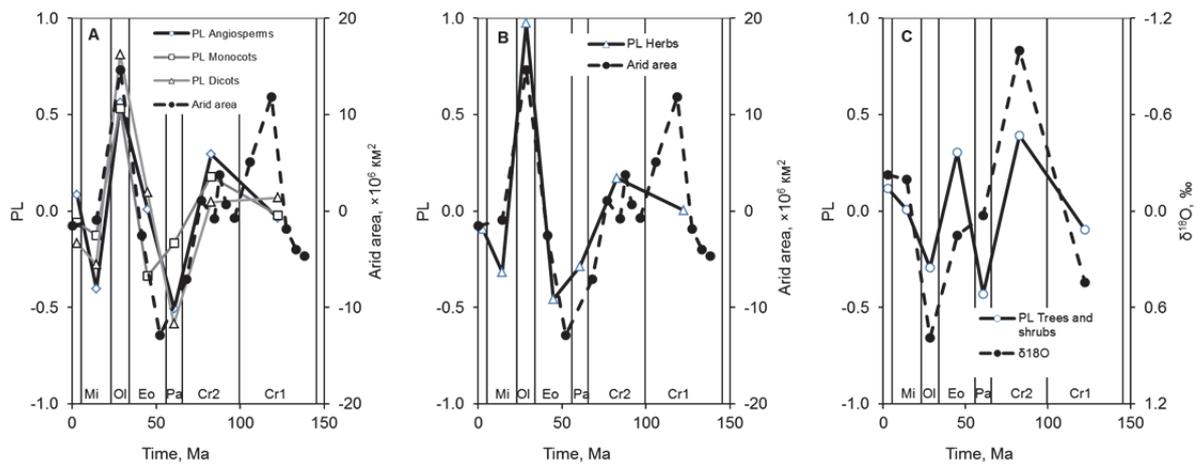

Fig. 8. The ploidy levels (PL) and the climate (A, B – area of arid territories; C – $\delta^{18}O$) vs. time. Data detrended.

The mentioned effects in Fig. 6–8 and other relations are shown as correlations coefficients in Table 4. The nDNAc in most of the plant groups was closely related with the activity of global hydrological cycle (in terms of arid areas occurrence) and the global temperature. Global cooling and climate aridization was followed by the significant increase of 2C. The SCN rather than 2n was related with the climatic changes, mainly the global temperature (Table 4). Temperature decrease in the Cretaceous–Cenozoic was followed by decrease of SCN. The PL was related mainly with the occurrence of arid areas. Climate

Table 4. Correlation coefficients between environment (arid area and global temperature) and nuclear DNA content (2C), numbers of chromosomes (2n), and ploidy levels (PL) in different groups of angiosperms (correlation coefficients are shown for log–transformed and detrended data)

| Plant group | Arid area | | | Global temperature | | |
|---|---|---|---|---|---|---|
| | 2C | 2n** | PL | 2C | 2n** | PL |
| Angiosperms | **0.81*** | –0.49 (–0.70) | 0.69 | **–0.83*** | 0.49 (**0.89***) | –0.60 |
| Monocots | 0.64 | **0.88*** (**0.82***) | **0.96*** | 0.37 | –0.48 (–0.59) | –0.69 |
| Dicots | **0.93*** | –0.64 (–0.73) | **0.81*** | **–0.75*** | 0.69 (**0.87***) | –0.49 |
| Trees and shrubs | **0.92*** | –0.64 (**–0.80***) | **–0.85*** | **–0.77*** | **0.98*** (**0.90***) | 0.66 |
| Herbs | 0.12 | –0.31 (–0.21) | **0.97*** | **–0.82*** | 0.09 (**0.86***) | –0.68 |

*Significant correlation coefficients (P<0.05)
** In parentheses are shown the correlation coefficients for specific chromosomes number (SCN=2n/2C)

aridization was followed by increase of PL in all plant groups not only in angiosperms as whole.

## THE NUCLEAR DNA CONTENT AND THE LEAF STRUCTURAL–FUNCTIONAL TRAITS

The nDNAc and some elements of the structural–functional organization of the plants have more or less equally changed in the geological past (FIG. 9). The index of plant water relations complexity ($I_{compl}$), an increase of which indicates increasing unfavorable environmental conditions (especially availability of the soil moisture) (Sheremet'ev, 2005), was varied not only due to changes in climate (Sheremet'ev and Gamalei, 2012) but also in accordance with changes of nDNAc (Fig. 9A).

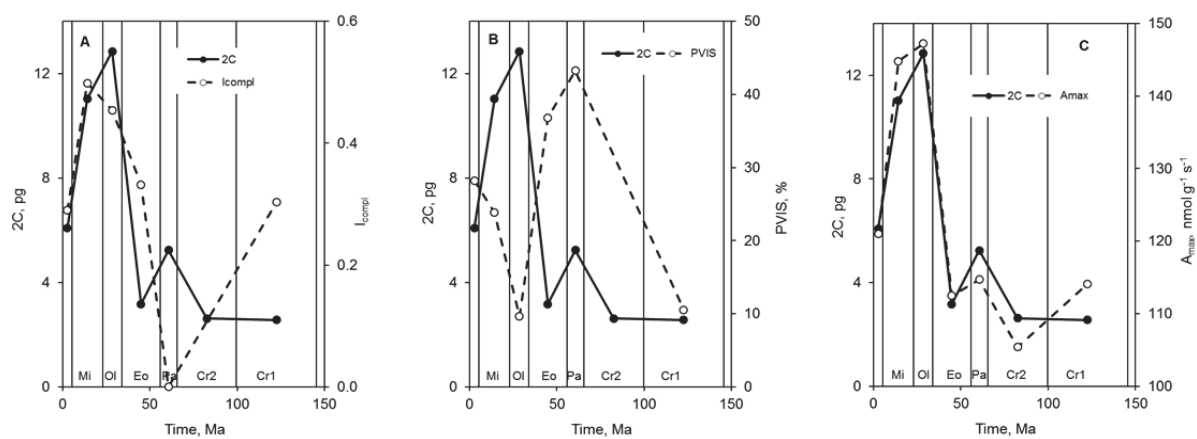

Fig. 9. The nuclear DNA content and the leaf structural-functional traits vs. time. A – index of the plant water relations complexity ($I_{compl}$), B – partial volume of intercellular spaces (PVIS), C – photosynthetic capacity ($A_{max}$) (A, B – after processing of data set by Sheremet'ev, 2005; C – after processing of data set by Wright et al., 2004).

Partial volume of intercellular spaces in leaves (PVIS) (lower values of this trait indicate strengthening of the xeromorphic organization of plants) was minimal in the most arid epochs of the Cretaceous–Cenozoic (Early Cretaceous, Oligocene, Miocene) and was maximal in the more humid epochs (Paleocene, Eocene) (Fig. 9B).

Photosynthetic capacity ($A_{max}$) of plants (data set by Wright et al., 2004) was closely correlated with the nDNAc (Fig. 9C) during Cretaceous–Cenozoic.

DISCUSSION

The available literatures discuss two options for the possible consequences of the genome increase: 1) an increase in functional capacity, 2) widening of the adaptive potentials of the species (Grime et al., 1985; Vinogradov, 2001, 2003; Gregory, 2001, 2002).
Structural or functional capacities of plants can be evaluated using the ratio of any trait value to 2C (structural or functional genome efficiency). For example, a comparison of plots of 2C and $A_{max}$ (Fig. 9C) shows their complete agreement (r = 0.975, P < 0.05) and seems to indicate that with the increase of nDNAc in the Oligocene–Miocene the functional capacities of the plants were growing. However, after calculating the ratio $A_{max}/2C$ it is clear that the functional efficiency of the genome ($A_{max}/2C$) was reduced (Fig. 10A). In other words, the growth of photosynthetic capacity was insufficient to maintain functional efficiency at the original level. The decrease of ratio $A_{max}/2C$ in the Cretaceous–Cenozoic was closely correlated with the decrease in the global temperature (Fig. 10A) (r = 0.813, P < 0.05). It seems that this global temperature trend was the cause of insufficient increase in $A_{max}$ in response to increased nDNAc.

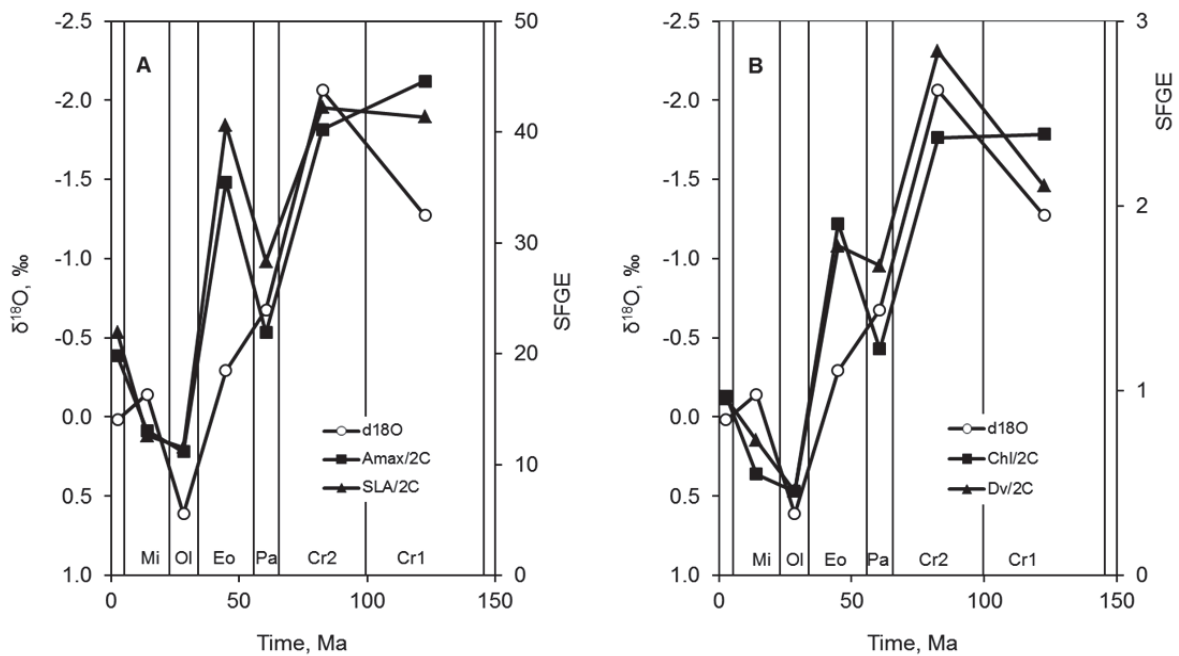

Fig. 10. The structural-functional genome efficiency (SFGE) (A – $A_{max}/2C$, SLA/2C; B – Chl/2C, Dv/2C) and the $\delta^{18}O$ vs. time. $A_{max}$ – photosynthetic capacity (after processing of data set by Wright et al., 2004), SLA specific leaf area (after processing of data set by Wright et al., 2004), Chl – chlorophyll content (after processing of data set by Lubimenko, 1916), Dv – vein density (after processing of data set by Brodribb, Feild, 2010 and Field et al., 2011).

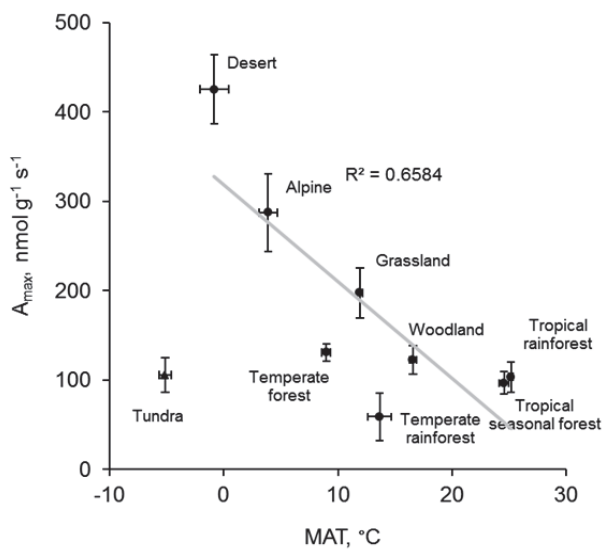

Fig. 11. The photosynthetic capacity ($A_{max}$) in different biomes (after processing of data set by Wright et al., 2004) (MAT – mean annual temperature).

There are several studies on the changes of $A_{max}$ on the latitudinal gradient, i.e. on the gradient of mean annual temperature (MAT). According to Woodward and Smith (1994) the highest values of the $A_{max}$ are typical for the tropical flora (20–30 $\mu mol \cdot m^{-2} \cdot c^{-1}$), and the smallest values – for the Arctic flora (3–7 $\mu mol \cdot m^{-2} \cdot c^{-1}$). According to other sources, the reduction of $A_{max}$ can be traced in the direction from the drought deciduous forests and temperate grasslands to tropical savanna and tropical rainforest (about from 26 to 16 $\mu mol \cdot m^{-2} \cdot c^{-1}$) (Schulze et al., 1994). Processing of the database GLOPNET (see Wright et al., 2004) shows that $A_{max}$ of angiosperms (n = 723) have increased with decreasing MAT (Fig. 11, r = 0.811, P < 0.05) (plants of the tundra are exceptions).

Despite of this, due to the rapid increase of 2C (see Bennett, 1976, 1987; Bennett and Leitch, 2005a), the functional efficiency of the genome (the rate of function per pg of DNA) should decrease in the direction from the tropical to the Arctic floras. In the Cretaceous–Cenozoic, a decrease in functional efficiency of the genome in parallel to decrease of the global temperature also occurred (Fig. 10A). This is related not only to $A_{max}$, but also to other traits: to specific leaf area (processing of data set by Wright et al., 2004), vein density (processing of data set by Brodribb and Feild, 2010 and Field et al., 2011), chlorophyll content (processing of data set by Lubimenko, 1916) (Fig. 10).

Thus, the growth of the genome in the Cenozoic did not lead to the intensification of functions, but rather led to the expansion of the adaptive capacity of species. The high content of DNA is a compensatory character which reflects the degree of defectiveness of the climate or soil conditions for plants. A genome's increase induced by an environmental stress expands resources of adaptive variations. Apparently correlation of the genome size with an adaptive potential is a peculiar property not only of vascular plants (Bennett and Leitch, 2005a, b; Knight et al., 2005), but also of other eukaryotes (Vinogradov, 1995, 1997, 1999; Gregory, 2001, 2002).

Growth of nuclear DNA content can be considered as one of the effective tools of an adaptogenesis. Herbaceous plants of the Neogene possessed, an order of magnitude larger genome size, and had a much broader ecological range than woody plants of the Paleogene. Herbs and shrubs turned out to be less specialized than the trees. They were able to grow in those climates, where the trees had no place. Herbal biomes ousted the forests over large areas in Neogene due to less demands to the environment. If the general trend of climate changes can be defined as a stepwise cooling, then at each stage, the growth of the genome size and the partial reset of the structural and functional specialization can be predicted.

ACKNOWLEDGEMENTS

This work was supported by a research grant from "The Russian Foundation for Basic Research" (10–04–01165–a).